\documentclass[a4paper]{jpconf}
\usepackage{graphicx}
\usepackage{epsfig}
\usepackage{amsfonts,bm}
\usepackage{amsfonts,amssymb,amsmath}
\usepackage{slashed}
\usepackage{txfonts}
\usepackage{lmodern}
\usepackage{exscale}
\usepackage{cite}
\usepackage[T1]{fontenc}
\usepackage[utf8]{inputenc}

\newcommand{\Dmn}{\Delta_{\mu\nu}}
\newcommand{\Pmn}{\Pi_{\mu\nu}}
\newcommand{\Del}{\Delta^{-1}}
\newcommand{\Vm}{V_{\mu\alpha\beta}}
\newcommand{\J}{J}

\DeclareMathOperator{\D}{\mathcal{D}}

\begin{document}
\title{Mass generation and the problem of seagull divergences}

\author{C. T. Figueiredo and A. C. Aguilar}

\address{University of Campinas - UNICAMP, 
Institute of Physics ``Gleb Wataghin'', \\
13083-859 Campinas, SP, Brazil}

\ead{ctfigue@ifi.unicamp.br}

\begin{abstract}
The gluon mass generation is a purely non-perturbative effect, and the natural framework to study it in the continuum are the Schwinger-Dyson equations (SDEs) of the theory. At the level of the SDEs the generation of such a mass is associated with the existence of infrared finite solutions for the gluon propagator. From the theoretical point of view, the dynamical gluon mass generation has been traditionally plagued with seagull divergences. In this work, we will review how such divergences can be eliminated completely by virtue of a characteristic identity, valid in dimensional regularization. As a pedagogical example, we will first discuss in the context of scalar QED how it is possible to eliminate all seagull divergences, by triggering the aforementioned special identity, which enforces the masslessness of the photon. Then, we will discuss what happens in QCD and present an Ansatz for the three gluon vertex, which completely eliminates all seagull divergences and at same time allows for the possibility of a dynamical gluon mass generation.
\end{abstract}

\section{Introduction}
\vspace{0.5cm}

The dynamical gluon mass generation has
been the subject of numerous studies in the past few years~\cite{Cornwall:1981zr,Aguilar:2006gr,Aguilar:2008xm,Binosi:2009qm,Aguilar:2009ke,Binosi:2012sj}. Being a non-perturbative effect, 
the most adequate formalism to study it in the continuum are the QCD Schwinger-Dyson equations (SDEs). Specifically, we will work in the framework provided by the synthesis of the pinch technique (PT)~\cite{Cornwall:1981zr,Aguilar:2008xm,Binosi:2009qm} with the background field method (BFM)~\cite{Abbott:1980hw}, known in the literature as the PT-BFM scheme.

The most crucial theoretical ingredient for obtaining out of the SDEs an infrared finite gluon propagator, without
interfering with the gauge invariance of the theory, is the existence of a set of special vertices endowed with massless poles. These poles
act as composite, longitudinally coupled Nambu-Goldstone
bosons, which at the same time trigger the Schwinger mechanism in QCD 
and  maintain the gauge invariance of the theory. Specifically, the form of the Ward identities (WIs) and the Slavnov-Taylor identities (STIs) are kept intact before and after the  dynamical generation of a gluon mass. It is interesting to mention that a series of secondary effects related to the  
existence of a gluon mass can be observed in the QCD dynamics~\cite{Aguilar:2001zy}.  For example,  it is expected that  the QCD effective charge ``freezes'' at a finite value in the infrared due the presence of the momentum dependent gluon mass in the argument of the leading logarithm~\cite{Cornwall:1981zr,Aguilar:2009nf}.

However, one of the QCD long-standing problems is how  to generate a  dynamical gluon mass without the appearance of seagull divergences,
\textit{i.e.} divergent integrals of the type $\int_k\Delta(k)$ and $\int_k k^2\Delta^2(k)$.
 Here, we will follow the same steps presented in the Ref.~\cite{Aguilar:2009ke} and  we will review  how these divergences can be eliminated by virtue of a special identity, valid in dimensional regularization. 
 
These proceedings is organized as follows. In Section 2, we present the main ingredients and the crucial relations necessary for the generation of a dynamical gluon mass through the Schwinger mechanism. We also briefly mention how the seagull divergences come into play. In Section 3, we explore a didactic example, within the context of scalar QED, showing how it is possible to  eliminate  all seagull divergences
by triggering the aforementioned characteristic identity. 
In addition, we discuss how this  identity is instrumental for ensuring the masslessness of the photon in the absence of bound-state poles, {\it i.e.} when the Schwinger mechanism is not triggered. In Section 4, we put together all theoretical ingredients introduced  
in the the previous sections, in order to construct an Ansatz for the QCD three gluon vertex. The aforementioned Ansatz  completely eliminates all seagull divergences and at the same time allows for the possibility of a dynamical gluon mass generation. Finally, in Section 5 we summarize our conclusions.

\section{Gluon Mass Generation and Seagull Divergences}
\vspace{0.5cm}

The gluon propagator, $\Delta_{\mu\nu}(q)$, can be written in $R_\xi$-type gauges as
\begin{equation}
\Dmn (q)=-i\left[P_{\mu\nu}(q)\Delta(q^2) +\xi\,\frac{q_\mu q_\nu}{q^4}\right] \,,
\label{eq:2.1}
\end{equation}
where $\xi$ is the gauge fixing parameter and $P_{\mu\nu}(q)$ is the transverse projector, given by
\begin{equation}
P_{\mu\nu}(q)=g_{\mu\nu}-\frac{q_\nu q_\mu}{q^2} \,.
\label{eq:2.2}
\end{equation}
The scalar factor $\Delta(q^2)$ can be obtained from 
\begin{equation}
\Del(q^2)=q^2+i\Pi(q^2) \,,
\label{eq:2.3}
\end{equation}
where $\Pi_{\mu\nu}(q) =P_{\mu\nu}(q)\,\Pi(q^2)$ is the gluon self-energy. 
It is interesting to mention that, a long time ago, Schwinger pointed out that if the theory has sufficient
strength  to produce a vacuum polarization of the gauge bosons with a pole at zero momentum transfer, 
the vector meson becomes massive, without breaking gauge symmetry~\cite{Schwinger:1962tp}. 
The dimensionless vacuum polarization, $\boldsymbol \Pi(q^2)$, is usually defined as $\Pi(q^2)=q^2\boldsymbol\Pi(q^2)$. Thus, it is clear that if $\boldsymbol\Pi(q^2)$ has a pole at $q^2=0$ with positive residue $\mu^2$, then
\begin{equation}
\Del(q^2)=q^2+\mu^2 \,,
\label{eq:2.5}
\end{equation}
in Euclidean space.
Since $\Del(0)=\mu^2$, the gluon becomes massive, even though no mass term is permitted at the level of the fundamental Lagrangian.
We note that the existence of this required pole for $\boldsymbol\Pi(q^2)$ occurs due to purely dynamical reasons,
in such a way that the dynamically generated mass for the vector meson does not interfere with gauge 
invariance~\cite{Jackiw:1973,Cornwall:1973,Eichten:1974}.

In order to obtain the momentum dependence of the  dynamical gluon mass, 
we must study the SDE that governs the gluon self-energy. 
The Schwinger mechanism is integrated into the SDE through the three-gluon vertex. However, despite the use of a three-gluon vertex containing massless poles and satisfying the correct WI, the value of the gluon propagator at zero momentum is expressed in terms of formally divergent seagull-type 
contributions of the form 
\begin{equation}
\Del(0)=c_1\int_k{\Delta(k)}+c_2\int_k{k^2\Delta^2(k)} \,.
\label{eq:2.7}
\end{equation}
where $\int_k\equiv\mu^{2\epsilon}(2\pi)^{-d}\int{d^dk}$ and $d=4-\epsilon$, in dimensional regularization (DR).
 
The downside of the presence of such divergent integrals is that they need to be regularized. It turns out that the regularization procedure  employed in the literature furnishes finite but scheme-dependent gluon 
masses~\cite{Aguilar:2008xm}.

Therefore,  the  full realization of gluon mass generation preserving all symmetries of the theory is an extremely subtle task. Basically, it demands that our Ansatz for the three gluon vertex has the following simultaneous features (i) contains massless poles, (ii) respect the corresponding WIs, and (iii)  preserves the cancellation that eliminates all seagull divergences of the type of Eq.~(\ref{eq:2.7}).

\section{Scalar QED} 
\vspace{0.5cm}

\begin{figure}[ht]
\begin{center}
\includegraphics[scale=0.55]{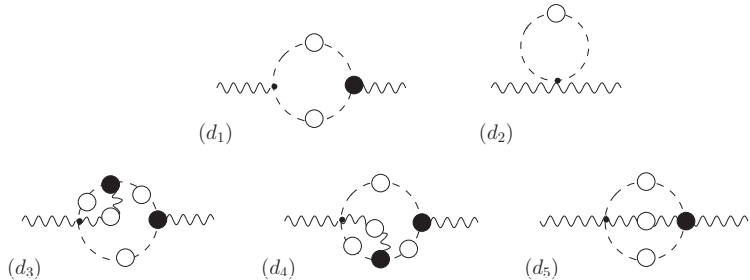}
\caption{Diagrams contributing to the SDE for the photon self energy in scalar QED~\cite{Aguilar:2009ke}.}
\label{fig1}
\end{center}
\end{figure}

Let us first analyze some of the basic issues regarding the appearance and cancellation of seagull divergent integrals in the context of scalar QED, which has a much simpler structure than QCD. 
We start by showing in  Fig.~\ref{fig1} the diagrams contributing to the SDE for the photon of scalar QED. 

Using the WIs satisfied by the vertices appearing in Fig.~\ref{fig1},  it is straightforward to demonstrate that the SDE may be truncated ``loop-wise'' ~\cite{Aguilar:2009ke}, 
 so that 
\begin{eqnarray}
q^\mu\Pmn (q)\left|_{\left[(d_1)+(d_2)\right]} \right.&=&0\,, \nonumber \\
q^\mu\Pmn (q)\left|_{\left[(d_3)+(d_4)+(d_5)\right]}\right.&=&0 \,.
\end{eqnarray}
 
\begin{figure}[hc]
\begin{center}
\includegraphics[scale=0.7]{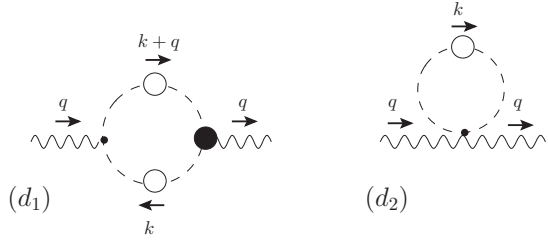}
\caption{The ``one-loop dressed'' contribution to the SDE for the photon self-energy in scalar QED.}
\label{fig2}
\end{center}
\end{figure}

In the Fig.~\ref{fig2} we show the terms contributing to the SDE for the photon self-energy at the ``one-loop dressed'' level, which can be written as
\begin{equation}
\Pmn(q)=e^2\int_k\Gamma_\mu^{(0)}\mathcal{D}(k)\D(k+q)\Gamma_\nu +e^2\int_k\Gamma_{\mu\nu}^{(0)}\D(k) \,,
\label{eq:3.1}
\end{equation}
where $\D(k)$ is the full propagator of the scalar field and $\Gamma_\nu$ is the full photon-scalar vertex, 
whose tree-level expression is
\begin{equation}
\Gamma_\nu^{(0)}=-i(2k+q)_\mu \,.
\label{eq:8}
\end{equation}
In addition, the bare quadrilinear photon-scalar vertex appearing in the diagram $(d_2)$ is expressed as $\Gamma_{\mu\nu}^{(0)}=2ig_{\mu\nu}$.

The Abelian WI for the photon-scalar vertex $\Gamma_\mu$ reads
\begin{equation}
q^\mu\Gamma_\mu=\D^{-1}(k+q)-\D^{-1}(k) \,.
\label{eq:3.2}
\end{equation}
With the help of the WI, one can easily show that $q^\nu\Pmn(q)=0$.  Therefore, $\Pmn(q)$ is transverse 
and assumes the form $\Pmn(q)=\Pi(q^2)P_{\mu\nu}(q)$. Thus, 
it follows that the scalar $\Pi(q^2)$ is given by
\begin{equation}
\Pi(q^2)=\frac{-2ie^2}{d-1}\left[\int_k\D(k)\D(k+q)k^\mu\,\Gamma_\mu -d\int_k\D(k)\right] \,.
\label{eq:3.3}
\end{equation}

In addition,  we can write the one-loop expression for $\Pi(q^2)$ as being
\begin{equation}
\Pi^{(1)}(q^2)=\frac{-ie^2}{d-1}\left[\int_k(4k^2-q^2)\D_0(k)\D_0(k+q) -2d\int_k\D_0(k)\right] \,,
\label{eq:3.4}
\end{equation}
where $\D_0(k)=(k^2-m^2)^{-1}$.
Now, we take the limit $q\rightarrow 0$, which leads to
\begin{equation}
\Pi^{(1)}(0)=\frac{-4ie^2}{d-1}\left[\int_k k^2\D^2_0(k) -\frac{d}{2}\int_k\D_0(k)\right] \,.
\label{eq:3.5}
\end{equation}
Using spherical coordinates in the Euclidean space, it can be demonstrated that \cite{Aguilar:2009ke} 
%
%
%
\begin{equation}
\int_k k^2\,\frac{\partial\D_0(k)}{\partial k^2} =-\frac{d}{2}\int_k\D_0(k) \,.
\label{eq:3.8}
\end{equation}
Consequently, we have
$\Pi^{(1)}(0)=0$, as expected, given that the photon remains massless perturbatively.
It is important to mention that the same result can be obtained using the following exact identity in DR:
\begin{equation}
\int_k\frac{k^2}{(k^2-m^2)^2} =\frac{d}{2}\int_k\frac{1}{k^2-m^2} \,.
\label{eq:3.6}
\end{equation}
The equation above can be verified applying the standard integration rules of DR~\cite{Peskin:1995ev}.

Now, let us analyze the full case given by Eq.~(\ref{eq:3.1}). Evidently, in order to preserve the transversality of $\Pmn(q)$ any Ansatz for $\Gamma_\mu$ should satisfy the WI of Eq.~(\ref{eq:3.2}). A suitable form is the one proposed  by Ball and Chiu~\cite{Ball:1980} given by
%
\begin{eqnarray}
\Gamma_\mu =\frac{(2k+q)_\mu}{(k+q)^2-k^2}\left[\D^{-1}(k+q)-\D^{-1}(k)\right]   + A(k,q)\left[(k+q)\cdot q\,k_\mu -k\cdot q\,(k+q)_\mu\right] \,,
\label{eq:3.10}
\end{eqnarray}
%
where $A(k,q)$ is finite as $q\rightarrow 0$.


The term proportional to $A(k,q)$, when inserted into Eq.~(\ref{eq:3.3}), yields a contribution that vanishes as $q\rightarrow 0$. Since we 
are interested  in the limit of vanishing $q$, only the 
the first term of $\Gamma_\mu$ contributes,  producing 
\begin{equation}
\Pi(0)=\frac{4ie^2}{d-1}\left[\int_k k^2\,\frac{\partial\D(k)}{\partial k^2} +\frac{d}{2}\int_k\D(k)\right] \,,
\label{eq:3.13}
\end{equation}

Certainly, we must have $\Pi(0)=0$, since we have not applied Schwinger's mechanism, 
{\it i.e.}, no dynamical poles were introduced in the Ansatz of Eq.~(\ref{eq:3.10}). Therefore, we must have
\begin{equation}
\int_k k^2\,\frac{\partial\D(k)}{\partial k^2}=-\frac{d}{2}\int_k\D(k) \,,
\label{eq:3.14}
\end{equation}
which is the non-perturbative generalization of Eq.~(\ref{eq:3.8}). 
The demonstration of this identity follows along the same lines as that for Eq.~(\ref{eq:3.8}).

We can now try another Ansatz for $\Gamma_\mu$ with massless poles, which will not trigger anymore the Eq.~(\ref{eq:3.14}).
Let us consider the vertex given by
\begin{equation}
\Gamma_\mu=\Gamma_\mu^{(0)} +\frac{q_\mu}{q^2}\left[\Sigma(k+q)-\Sigma(k)\right] \,,
\label{eq:3.15}
\end{equation}
where $\Sigma(k)$ is the self-energy of the scalar field 
and $\D^{-1}(k)=k^2-m^2+\Sigma(k)$. Thus,
\begin{equation}
\Gamma_\mu=\left\{\Gamma_\mu^{(0)} -\frac{q_\mu}{q^2}\left[(k+q)^2-k^2\right]\right\} 
+\frac{q_\mu}{q^2}\left[\D^{-1}(k+q)-\D^{-1}(k)\right] \,,
\label{eq:3.16}
\end{equation}
which satisfies the WI of Eq.~(\ref{eq:3.2}), 
but contains massless poles. 
Substituting $\Gamma_\mu$ of Eq.~(\ref{eq:3.16}) into Eq.~(\ref{eq:3.3}), one finds
\begin{equation}
\Pi(0)=ie^2\left[2\int_k\D(k)-\int_k k^2\D^2(k)\right] \,,
\label{eq:3.19}
\end{equation}
which possess the general form of Eq.~(\ref{eq:2.7}). 
Hence, we end up with a non-zero $\Delta^{-1}(0)$, but expressed in terms of seagull divergences.

In the next section, then, we will look for an Ansatz, in the case of a pure Yang-Mills theory, 
that adds non-perturbative pole terms to the vertex of Eq.~(\ref{eq:3.10}),
but also preserves the cancellation implemented by Eq.~(\ref{eq:3.14}), 
in order to obtain a finite $\Delta^{-1}(0)$.


\section{Finite Gluon Mass Generation}
\vspace{0.5cm}

\begin{figure}[h!]
\begin{center}
\includegraphics[scale=0.7]{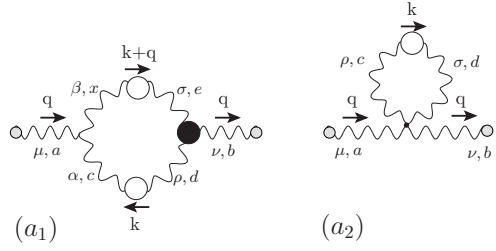}
\caption{The ``one-loop dressed'' gluonic graphs for the SDE of the (background) gluon self-energy.}
\label{fig3}
\end{center}
\end{figure}

\begin{figure}[h!]
\vspace{1.0cm}
\begin{minipage}[b]{0.3\linewidth}
\begin{center}
\includegraphics[scale=0.6]{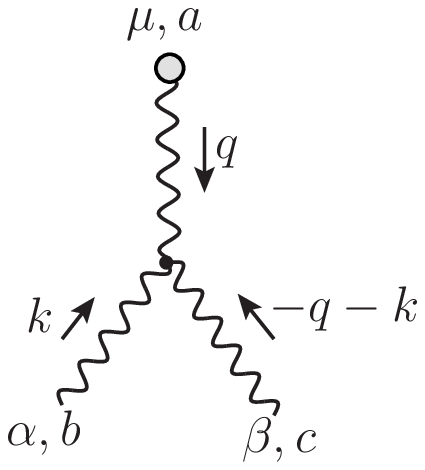}
\end{center}
\end{minipage}
\vspace{0.6cm}
\hspace{0.2cm}
\begin{minipage}[b]{0.7\linewidth}
\begin{center}
\begin{eqnarray}
g\,f^{abc}\left[(2k+q)_\mu g_{\alpha\beta}+2q_\beta g_{\mu\alpha}-2q_\alpha g_{\mu\beta}\right] \,,
\nonumber
\end{eqnarray}
\end{center}
\end{minipage}
\begin{minipage}[b]{0.3\linewidth}
\begin{center}
\includegraphics[scale=0.55]{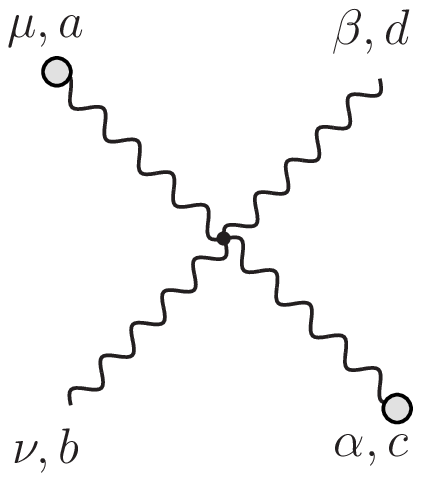}
\end{center}
\end{minipage}
\hspace{0.4cm}
\begin{minipage}[b]{0.7\linewidth}
\begin{center}
\begin{eqnarray}
-ig^2\,\left[f^{abx}f^{xcd}(g_{\mu\alpha}g_{\nu\beta}-g_{\mu\beta}g_{\nu\alpha}+g_{\mu\nu}g_{\alpha\beta})\right. \nonumber \\
\left.+f^{adx}f^{xbc}(g_{\mu\nu}g_{\alpha\beta}-g_{\mu\alpha}g_{\nu\beta}+g_{\mu\beta}g_{\nu\alpha}) \right. \nonumber \\
\left.+f^{acx}f^{xbd}(g_{\mu\nu}g_{\alpha\beta}-g_{\mu\beta}g_{\nu\alpha})\right]  \,. \nonumber
\end{eqnarray}
\end{center}
\end{minipage}
\caption{\label{fig4} The trilinear and quadrilinear gluon vertices in the Feynman gauge of PT-BFM~\cite{Aguilar:2009ke}.}
\end{figure}

Let us now turn to the case of pure QCD and analyze the SDE of the background gluon propagator within the PT-BFM framework. The advantage of working with the PT-BFM formalism, it is that  the transversality of the gluon self-energy  is preserved at every step,  allowing for a gauge-invariant truncation of the SD series~\cite{Binosi:2009qm, Aguilar:2006gr}.
Here, we will study only the ``one-loop dressed'' gluonic part of the SDE diagrammatically represented in the Fig.~\ref{fig3}. In addition, the  Feynman rules for the two  tree-level vertices appearing in the PT-BFM formulation 
are presented in Fig.~\ref{fig4}.

To reduce the algebraic complexity of the problem, we set $\Delta_{\alpha\beta}\rightarrow -ig_{\alpha\beta}\Delta$.
In order to be consistent and maintain the transversality of the gluon self-energy, $\widehat{\Pi}_{\mu\nu}(q)$, 
we write the WI satisfied by $\widetilde{\Gamma}_{\nu\alpha\beta}$ as
\begin{equation}
q^\nu\widetilde{\Gamma}_{\nu\alpha\beta} =\left[\Del(k+q)-\Del(k)\right]g_{\alpha\beta} \,.
\label{eq:4.3}
\end{equation} 
Then, it can be demonstrated that the mathematical expression for the SDE of Fig.~\ref{fig3} is
\begin{equation}
\widehat{\Delta}^{-1}(q) =q^2+\frac{iq^2C_A}{2(d-1)}\left[\int_k\widetilde{\Gamma}^{(0)}_{\mu\alpha\beta}\Delta(k)\Delta(k+q)\widetilde{\Gamma}^{\mu\alpha\beta} +2d^2\int_k\Delta(k)\right] \,,
\label{eq:4.4}
\end{equation}
where $C_A$ is the Casimir eigenvalue of the adjoint representation, $\widetilde{\Gamma}^{(0)}_{\mu\alpha\beta}(q,k,-k-q)$ is the bare BFM three-gluon vertex in the Feynman gauge (see Fig.~\ref{fig4}), and $\widetilde{\Gamma}_{\mu\alpha\beta}$ represents its fully-dressed version.

\vspace{0.5cm}
The function $\widehat{\Delta}(q)$ appearing on the lhs of Eq.~(\ref{eq:4.4}) is related to the conventional $\Delta(q)$ through 
\begin{equation}
\widehat{\Delta}(q)[1+G(q^2)]^2=\Delta(q) \,,
\label{eq:23}
\end{equation}
where $G(q^2)$ is the auxiliary two-point function of the PT-BFM framework~\cite{Binosi:2009qm,Binosi:2002ez,Binosi:2002ft,Binosi:2003rr}.
To simplify our work, 
we set $G(q^2)=0$, which does not alter the essential features we intend to check. 
Therefore, we assume $\widehat{\Delta}(q)=\Delta(q)$, so that we can suppress the ``hats'' from now on.

We can now write $\Del(q)$ (in Minkowski space) as
\begin{equation}
\Del(q)=q^2\J(q)-\widetilde{m}^2(q) \,.
\label{eq:4.5}
\end{equation}
Then, an adequate Ansatz for $\widetilde{\Gamma}_{\nu\alpha\beta}$ is
\begin{equation}
i\widetilde{\Gamma}_{\nu\alpha\beta} =\left[\frac{(k+q)^2\J(k+q)-k^2\J(k)}{(k+q)^2-k^2}\right]\widetilde{\Gamma}_{\nu\alpha\beta}^{(0)} +\Vm \,,
\label{eq:4.6}
\end{equation}
where $\Vm$ contains the non-perturbative contributions due to bound-state poles associated with the Schwinger mechanism.
Note that, in order for $\widetilde{\Gamma}_{\nu\alpha\beta}$ to satisfy the WI of Eq.~(\ref{eq:4.3}), we should have
\begin{equation}
q^\mu\Vm=\left[\widetilde{m}^2(k)-\widetilde{m}^2(k+q)\right]g_{\alpha\beta} \,.
\label{eq:4.7}
\end{equation}

The Ansatz of Eq.~(\ref{eq:4.6}) is somewhat analogous to that of Eq.~(\ref{eq:3.10}). The first term presents the correct structure 
for obtaining the derivative term of Eq.~(\ref{eq:3.14}), once we insert it into Eq.~(\ref{eq:4.4}). In addition, the rhs of Eq.~(\ref{eq:3.14}) is obtained from the second term of the rhs of Eq.~(\ref{eq:4.4}).
%
Therefore, we can eliminate the seagull divergence, and $\Vm$ can produce an IR-finite gluon propagator, $\Del(0)=-\widetilde{m}^2(0)$.

Moreover, $\Vm$ can be decomposed in a transverse and a longitudinal piece, such that  
\begin{equation}
\Vm=\Vm^l+\Vm^t \,,
\label{eq:4.9}
\end{equation}
where 
\begin{equation}
\Vm^l=\frac{q_\mu}{q^2}\left[\widetilde{m}^2(k)-\widetilde{m}^2(k+q)\right]g_{\alpha\beta} \,,
\label{eq:4.10}
\end{equation}
and 
\begin{equation}
q^\mu\Vm^t=0 \,.
\label{eq:4.11}
\end{equation}
Thus, we can write the vertex of Eq.~(\ref{eq:4.6}) as
\begin{equation}
i\widetilde{\Gamma}_{\mu\alpha\beta} =\left[\frac{\Del(k+q)-\Del(k)}{(k+q)^2-k^2}\right]\widetilde{\Gamma}_{\mu\alpha\beta}^{(0)} +\overline{V}_{\mu\alpha\beta} \,,
\label{eq:4.12}
\end{equation}
with
\begin{equation}
\overline{V}_{\mu\alpha\beta}= \Vm +(2k+q)_\mu\left[\frac{\widetilde{m}^2(k+q)-\widetilde{m}^2(k)}{(k+q)^2-k^2}\right]g_{\alpha\beta} \,.
\label{eq:4.13}
\end{equation}
This last form of $\widetilde{\Gamma}_{\mu\alpha\beta}$ allows for the immediate use of the identity of Eq.~(\ref{eq:3.14}).
Hence, the final non-perturbative effective vertex $\overline{V}_{\mu\alpha\beta}$ is transverse:
\begin{equation}
q^\mu\overline{V}_{\mu\alpha\beta} =0 \,.
\label{eq:4.15}
\end{equation}

Therefore in Eq.~(\ref{eq:4.12}), we were able to propose an Ansatz for the vertex with three important characteristics:
\begin{enumerate}
	\item It satisfies the WIs of Eq.~(\ref{eq:4.3}), ensuring the transversality of the gluon self-energy;
	\item The pole term contained in $\Vm$ allows for a non-vanishing $\Del(0)$;
	\item It triggers Eq.~(\ref{eq:3.14}), which cancels out all the the seagull integrals.
\end{enumerate}
Consequently, the resulting $\Del(0)$ is both non-vanishing and finite, according to our goal.

\section{Conclusions}
\vspace{0.5cm}

In this work, we have demonstrated the possibility of eliminating the problem of seagull divergences 
in the generation of a dynamical gluon mass by means of an insightful Ansatz for the three gluon vertex.
This Ansatz is given in Eq.~(\ref{eq:4.12}) and its form mimics that of the Ball and Chiu vertex in the case of scalar QED.
Specifically, the Ansatz in question was chosen so that it would trigger the seagull identity of Eq.~(\ref{eq:3.14}), valid in dimensional regularization.
Therefore, this method allows for the dynamical generation of a non-vanishing but finite gluon mass free of seagull divergences. 

\ack{\vspace{0.5cm} We would like to thank the organizers of the  XIII  Hadron Physics  for the pleasant workshop.  The research of ACA is supported  by 
National Council for Scientific and
Technological Development-CNPq under the grant 306537/20
12-5 and project 473260/2012-3,
and by S\~ao Paulo Research Foundation - FAPESP through the project 2012/15643-1. The work of CTF is supported by FAPESP under the grant 2014/16247-8.


\section*{References}
\vspace{0.5cm}

\begin{thebibliography}{9}

\bibitem{Cornwall:1981zr} 
  J.~M.~Cornwall,
  Phys.\ Rev.\ D {\bf 26}, 1453 (1982).

\bibitem{Aguilar:2006gr} 
  A.~C.~Aguilar and J.~Papavassiliou,
  JHEP {\bf 0612}, 012 (2006).

\bibitem{Aguilar:2008xm} 
  A.~C.~Aguilar, D.~Binosi and J.~Papavassiliou,
  Phys.\ Rev.\ D {\bf 78}, 025010 (2008).



\bibitem{Binosi:2009qm} 
  D.~Binosi and J.~Papavassiliou,
  Phys.\ Rept.\  {\bf 479}, 1 (2009).
  
\bibitem{Aguilar:2009ke} 
  A.~C.~Aguilar and J.~Papavassiliou,
  Phys.\ Rev.\ D {\bf 81}, 034003 (2010).


\bibitem{Binosi:2012sj} 
  D.~Binosi, D.~Ibanez and J.~Papavassiliou,
  Phys.\ Rev.\ D {\bf 86}, 085033 (2012).



\bibitem{Abbott:1980hw} 
  L.~F.~Abbott,
  Nucl.\ Phys.\ B {\bf 185}, 189 (1981).


\bibitem{Aguilar:2001zy} 
  A.~C.~Aguilar, A.~Mihara and A.~A.~Natale,
  Phys.\ Rev.\ D {\bf 65}, 054011 (2002).


\bibitem{Aguilar:2009nf} 
  A.~C.~Aguilar, D.~Binosi, J.~Papavassiliou and J.~Rodriguez-Quintero,
  Phys.\ Rev.\ D {\bf 80}, 085018 (2009).

 
\bibitem{Schwinger:1962tp} 
  J.~S.~Schwinger,
  Phys.\ Rev.\  {\bf 128}, 2425 (1962).
  
  \bibitem{Jackiw:1973} 
  R.~Jackiw and K.~Johnson, 
  Phys.\ Rev.\ D {\bf 8}, 2386 (1973).
  
  \bibitem{Cornwall:1973} 
  J.~M.~Cornwall and R.~E.~Norton, 
  Phys.\ Rev.\ D {\bf 8}, 3338 (1973).
  
    \bibitem{Eichten:1974} 
  E.~Eichten and F.~Feinberg, 
  Phys.\ Rev.\ D {\bf 10}, 3254 (1974).
  

  
\bibitem{Peskin:1995ev} 
  M.~E.~Peskin and D.~V.~Schroeder,
  ``An Introduction to quantum field theory'',
  Reading, USA: Addison-Wesley (1995) 842 p.
  
\bibitem{Ball:1980}
J.~S.Ball and T.~W.~Chiu,
Phys.\ Rev.\ D {\bf 22}, 2542 (1980).




\bibitem{Binosi:2002ez} 
  D.~Binosi and J.~Papavassiliou,
  Phys.\ Rev.\ D {\bf 66}, 025024 (2002).


\bibitem{Binosi:2002ft} 
  D.~Binosi and J.~Papavassiliou,
  Phys.\ Rev.\ D {\bf 66}, 111901 (2002).

\bibitem{Binosi:2003rr} 
  D.~Binosi and J.~Papavassiliou,
  J.\ Phys.\ G {\bf 30}, 203 (2004).


\end{thebibliography}
\end{document}